\pdfoutput=1

\documentclass[fleqn]{an}

\usepackage{graphicx,times}
\usepackage{natbib}
\usepackage{amssymb,amsmath}
\usepackage{hyperref}
\bibpunct{(}{)}{;}{a}{}{,}

\begin{document}

\Pagespan{614}{620}
\Yearpublication{2017}
\Yearsubmission{2015}
\Month{05}
\Volume{338}
\Issue{5}
\DOI{10.1002/asna.201713152}

\title{Distinguishing between order and chaos in a simple barred galaxy model}

\author{Euaggelos E. Zotos\inst{}\fnmsep\thanks{Corresponding author:
  \email{evzotos@physics.auth.gr}\newline}}
\titlerunning{Order and chaos in a barred galaxy model}
\authorrunning{Euaggelos E. Zotos}
\institute{Department of Physics, School of Science, Aristotle University of Thessaloniki,
GR-541 24, Thessaloniki, Greece}

\received{2 December 2015}
\accepted{21 February 2017}
\publonline{18 April 2017}


\abstract{We use a simple dynamical model in order to investigate the regular or chaotic character of orbits in a barred galaxy with a central, spherically symmetric, dense nucleus and a flat disk. In particular, we explore how the total orbital energy influences the overall orbital structure of the system, by computing in each case the percentage of regular, sticky and chaotic orbits. In an attempt to distinguish safely and with certainty between ordered and chaotic motion, we apply the Smaller ALingment Index (SALI) as a chaos detector to extensive samples of orbits obtained by integrating numerically the basic equations of motion as well as the variational equations. We integrate large sets of initial conditions of orbits in several types of two dimensional planes for better understanding of the orbital properties. Our numerical calculations suggest, that the value of the energy has a huge impact on the percentages of the orbits, thus indicating that a rotating barred galaxy is indeed a very interesting stellar quantity.}

\maketitle

\section{Introduction}
\label{intro}

It is well known that the axial symmetry in galaxies is only a first approach. In essence, galaxies exhibit deviation from the axial symmetry, which can be very small or extended. In the latter category, we may include the case of barred galaxies. Observations indicate, that a large percentage of disk galaxies, about 65\%, shows bar-like formations \citep[e.g.,][]{Ee00,SRSS03}. Observations also show, that barred galaxies may display different characteristics. There are galaxies with a prominent barred structure and also galaxies with faint weak bars. Moreover, there are also barred galaxies with massive and less massive bulges. In some cases, the formation of the central bulges is caused by dynamical instabilities of the disk \citep{KK04}.

Over the last decades, a huge amount of studies has been devoted on understanding the orbital structure in barred galaxy models \citep[e.g.,][]{ABMP83,P84,CDFP90,A92,P96,KC96,OP98,PMM04}. The reader can find more information about the dynamics of barred galaxies in the reviews \citet{A84,CG89,SW93}. We would like to point out, that all the above-mentioned references on the dynamics of barred galaxies are exemplary rather than exhaustive. However, it is useful to briefly discuss some of the recent papers on this subject. \citet{SPA02a} conducted an extensive investigation regarding the stability and the morphology of both two dimensional (2D) and three dimensional (3D) periodic orbits in a fiducial model, representative of a barred galaxy. The work was continued in \citet{SPA02b}, where the influence of the system's parameters on the periodic orbits was revealed. Moreover, \citet{KP05} presented evidence that in two dimensional models with sufficiently large bar axial ratios, stable orbits having propeller shapes play a dominant role to the bar structure.

In the same vein, \citet{MA11} estimated the fraction of chaotic and regular orbits in both two and three dimensional potentials by computing several sets of initial conditions of orbits and studying how these fractions evolve when the energy and also basic parameter of the model, such as the mass, the size and the pattern speed of the bar, vary. Computing the statistical distributions of sums of position coordinates \citet{BMA12}, quantified weak and strong chaotic orbits in 2D and 3D barred galaxy models. A time-dependent barred galaxy model was utilized in \citet{MBS13} in order to explore the interplay between chaotic and regular behavior of star orbits when the parameters of the model evolve. In \citep{CP07,CZ10} we conducted an investigation regarding only the issue of regular and chaotic orbits in simple barred spiral potentials. Finally in \citet{JZ15} a new realistic model was introduced for describing the properties of a barred galaxy, while in \citet{JZ16} the orbital and escape dynamics of this new model have been revealed.

In an earlier paper \citep{CP04} the global and local motion in a barred galaxy model with a massive nucleus had been studied. It was revealed that low and high energy stars in the global model display chaotic motion. In the local model on the other hand, low energy stars display resonance phenomena while the chaotic phenomena, if any, are negligible. The work was continued in \citet{Z12c} where an oblate halo component was added in the total gravitational potential. The same global model was used in \citet{EP14} for investigating the dynamics in the close vicinity of and within the critical area of the effective galactic potential. For values of energy larger than the critical escape energy, the escape dynamics of the stars were reveled and the corresponding basins of escape were identified. The authors also showed that the particular effective potential can realistically represent the formation as well as the evolution of the twin spiral arms located at the two edges of the galactic bar. In this work we shall use the same model-potential in an attempt to determine how the total orbital energy influences the ordered or chaotic nature of orbits for energy levels below the escape energy.

The present paper is organized as follows: In Section \ref{galmod} we present a detailed description of the properties of the barred galaxy model. All the different computational methods used in order to determine the nature of orbits are described in Section \ref{cometh}. In the following Section, we explore how the total orbital energy influences the percentages of regular and chaotic orbits. Our article ends with Section \ref{disc}, where the conclusions and the discussions of this research are presented.

\section{Properties of the galactic model}
\label{galmod}

The total gravitational potential $\Phi(x,y)$ consists of three components: the central spherical component $\Phi_{\rm n}$, the bar potential $\Phi_{\rm b}$ and the disk component $\Phi_{\rm d}$.

The spherically symmetric nucleus is modeled by a Plummer potential \citep[e.g.,][]{BT08}
\begin{equation}
\Phi_{\rm n}(x,y) = - \frac{G M_{\rm n}}{\sqrt{x^2 + y^2 + c_{\rm n}^2}}.
\label{Vn}
\end{equation}
Here $G$ is the gravitational constant, while $M_{\rm n}$ and $c_{\rm n}$ are the mass and the scale length of the nucleus, respectively. This potential has been used successfully in the past in order to model and therefore interpret the effects of the central mass component in a galaxy \citep[see e.g.,][]{HN90,HPN93,Z12a,ZC13}. At this point, we should emphasize that Eq. (\ref{Vn}) is not intended to represent the potential of a black hole nor that of any other compact object, but just the potential of a dense and massive nucleus therefore, any relativistic effects are out of the scope of this work.

For the description of the properties of the bar we use the following anharmonic mass-model potential
\begin{equation}
\Phi_{\rm b}(x,y) = - \frac{G M_{\rm b}}{\sqrt{x^2 + b^2 y^2 + c_{\rm b}^2}},
\label{Vb}
\end{equation}
where $M_{\rm b}$ and $c_{\rm b}$ are the mass and the scale length of the bar, respectively, while $b$ is a additional parameter controlling the strength of the bar.

In order to model the galactic disk we also use a Plummer potential
\begin{equation}
\Phi_{\rm d}(x,y) = - \frac{G M_{\rm d}}{\sqrt{x^2 + y^2 + c_{\rm d}^2}},
\label{Vd}
\end{equation}
where $M_{\rm d}$ and $c_{\rm d}$ are the mass and the scale length of the disk, respectively.

We consider the case where the bar rotates counterclockwise at a constant angular velocity $\Omega_{\rm b}$. Therefore the corresponding effective potential is
\begin{equation}
\Phi_{\rm eff}(x,y) = \Phi(x,y) - \frac{1}{2}\Omega_{\rm b}^2 \left(x^2 + y^2 \right).
\label{Veff}
\end{equation}

In our study, we use the well-known system of galactic units, where the unit of length is 1 kpc, the unit of mass is $2.325 \times 10^7 {\rm M}_\odot$ and the unit of time is $0.9778 \times 10^8$ yr. The velocity units is 10 km s$^{-1}$, the unit of angular momentum (per unit mass) is 10 km kpc$^{-1}$ s$^{-1}$, while $G$ is equal to unity $(G = 1)$. The energy unit (per unit mass) is 100 km$^2$s$^{-2}$. In these units, the values of the involved parameters are: $M_{\rm n} = 400$ (corresponding to 9.3 $\times$ $10^{9}$ M$_{\odot}$), $c_{\rm n} = 0.25$, $M_{\rm b} = 3000$ (corresponding to 6.975 $\times$ $10^{10}$ M$_{\odot}$), $b = 2$, $c_{\rm b} = 1.5$, $M_{\rm d} = 9500$ (corresponding to 2.2 $\times$ $10^{11}$ M$_{\odot}$), $c_{\rm d} = 12$, and $\Omega_{\rm b} = 1.25$. This set of the values of the parameters defines the Standard Model (SM) and remains constant throughout the numerical calculations.

The basic equations of motion are
\begin{equation}
\ddot{x} = - \frac{\partial \Phi_{\rm eff}}{\partial x} + 2\Omega_{\rm b}\dot{y}, \ \ \
\ddot{y} = - \frac{\partial \Phi_{\rm eff}}{\partial y} - 2\Omega_{\rm b}\dot{x},
\label{eqmot}
\end{equation}
where the dot indicates derivative with respect to the time.

In the same vein, the equations describing the evolution of a deviation vector ${\bf{w}} = (\delta x, \delta y, \delta \dot{x}, \delta \dot{y})$ which joins the corresponding phase space points of two initially nearby orbits, needed for the calculation of standard chaos indicators (the SALI in our case) are given by the following variational equations
\begin{eqnarray}
\dot{(\delta x)} &=& \delta \dot{x}, \ \ \
\dot{(\delta y)} = \delta \dot{y}, \nonumber \\
(\dot{\delta \dot{x}}) &=&
- \frac{\partial^2 \Phi_{\rm eff}}{\partial x^2} \delta x
- \frac{\partial^2 \Phi_{\rm eff}}{\partial x \partial y}\delta y + 2\Omega_{\rm b} \delta \dot{y},
\nonumber \\
(\dot{\delta \dot{y}}) &=&
- \frac{\partial^2 \Phi_{\rm eff}}{\partial y \partial x} \delta x
- \frac{\partial^2 \Phi_{\rm eff}}{\partial y^2}\delta y - 2\Omega_{\rm b} \delta \dot{x}.
\label{vareq}
\end{eqnarray}

Consequently, the corresponding Hamiltonian to the effective potential given in Eq. (\ref{Veff}) reads
\begin{equation}
H(x,y,\dot{x},\dot{y}) = \frac{1}{2} \left(\dot{x}^2 + \dot{y}^2 \right) + \Phi_{\rm eff}(x,y) = E,
\label{ham}
\end{equation}
where $\dot{x}$ and $\dot{y}$ are the velocities, while $E$ is the numerical value of the energy integral, which is conserved. Thus, an orbit with a given value for it's Jacobi integral is restricted in its motion to regions in which $E \leq \Phi_{\rm eff}$, while all other regions are forbidden to the stars.

\section{Computational methods}
\label{cometh}

In order to investigate the orbital structure of the galaxy, we need to define samples of initial conditions of orbits whose properties (order or chaos) will be identified. For this purpose, we define for each value of the total orbital energy (all tested energy levels are below the critical escape energy), dense uniform grids of $1024 \times 1024$ initial conditions regularly distributed in the area allowed by the value of the energy $E$. Our investigation takes place both in both the configuration $(x,y)$ and the phase $(x,\dot{x})$ space, for a better understanding of the orbital content. Furthermore, the grids of initial conditions of orbits whose properties will be explored are defined as follows: For the configuration $(x,y)$ space we consider orbits with initial conditions $(x_0, y_0)$ with $\dot{x_0} = 0$, while the initial value of $\dot{y_0}$ is always obtained from the energy integral (\ref{ham}) as $\dot{y_0} = \dot{y}(x_0,y_0,\dot{x_0},E) > 0$. Similarly, for the phase $(x,\dot{x})$ space we consider orbits with initial conditions $(x_0, \dot{x_0})$ with $y_0 = 0$, while the value of $\dot{y_0}$ is again obtained from the energy integral.

The equations of motion (\ref{eqmot}) as well as the variational equations (\ref{vareq}) were integrated using a double precision Bulirsch-Stoer \verb!FORTRAN 77! algorithm \citep[e.g.,][]{PTVF92} with a small time step of order of $10^{-2}$, which is sufficient enough for the desired accuracy of our computations (i.e., our results practically do not change by halving the time step). Here we should emphasize, that our previous numerical experience suggests that the Bulirsch-Stoer integrator is both faster and more accurate than a double precision Runge-Kutta-Fehlberg algorithm of order 7 with Cash-Karp coefficient. Throughout all our computations, the energy energy integral (Eq. (\ref{ham})) was conserved better than one part in $10^{-12}$, although for most orbits it was better than one part in $10^{-13}$.

When studying the orbital structure of a dynamical system, knowing whether an orbit is regular or chaotic is an issue of significant importance. Over the years, several dynamical indicators have been developed in order to determine the nature of orbits. In our case, we chose to use the Smaller ALingment Index (SALI) method. The SALI \citep{S01} is undoubtedly a very fast, reliable and effective tool, which is defined as
\begin{equation}
\rm SALI(t) \equiv min(d_-, d_+),
\label{sali}
\end{equation}
where $d_- \equiv \| {\bf{w_1}}(t) - {\bf{w_2}}(t) \|$ and $d_+ \equiv \| {\bf{w_1}}(t) + {\bf{w_2}}(t) \|$ are the alignments indices, while ${\bf{w_1}}(t)$ and ${\bf{w_2}}(t)$, are two deviations vectors which initially point in two random directions. For distinguishing between ordered and chaotic motion, all we have to do is to compute the SALI for a relatively short time interval of numerical integration $t_{\rm max}$. More precisely, we track simultaneously the time-evolution of the main orbit itself as well as the two deviation vectors ${\bf{w_1}}(t)$ and ${\bf{w_2}}(t)$ in order to compute the SALI. The variational equations (\ref{vareq}), as usual, are used for the evolution and computation of the deviation vectors.

\begin{figure}[!t]
\centering
\includegraphics[width=\hsize]{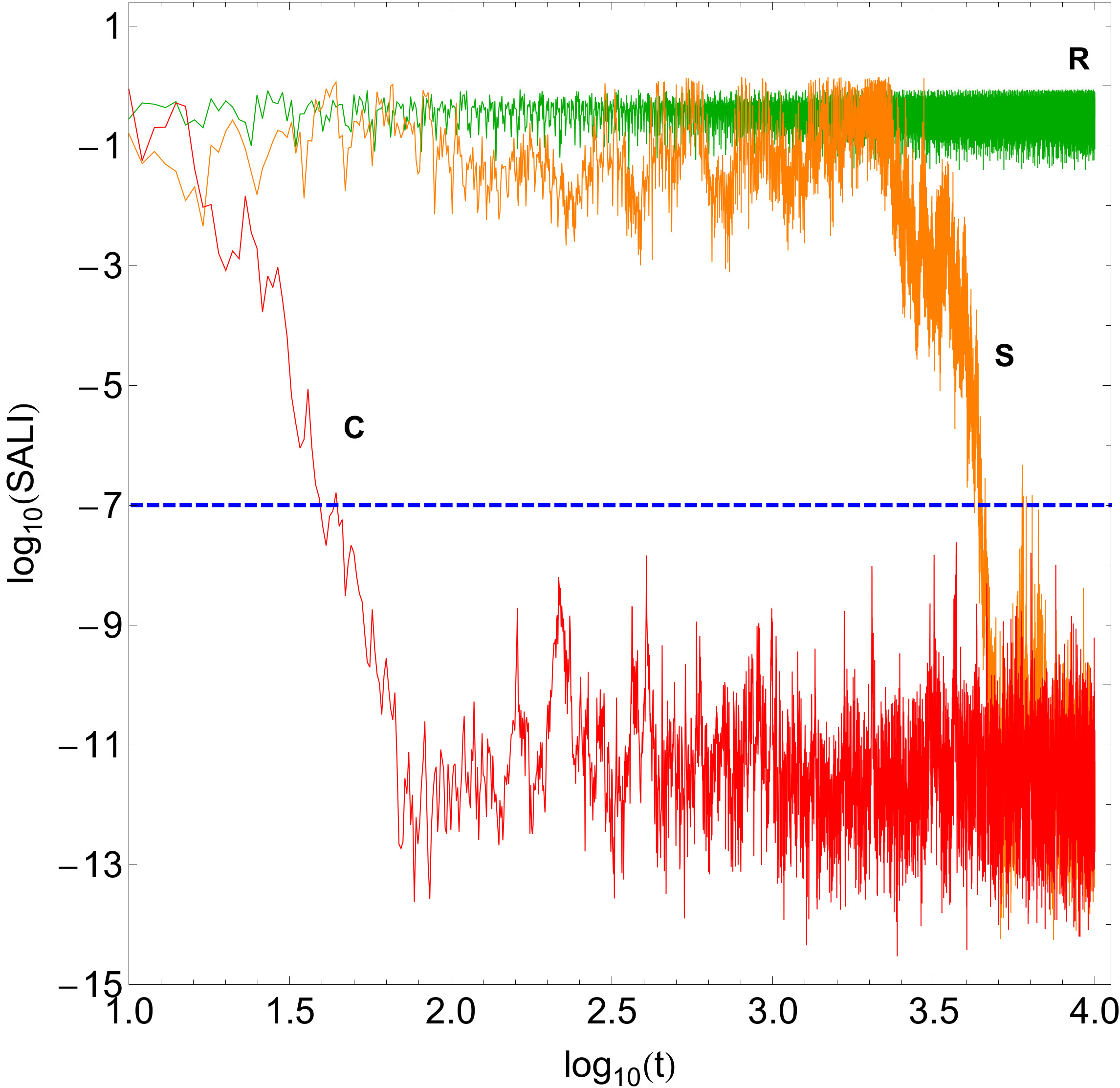}
\caption{Time-evolution of the SALI of a regular orbit (green color - R), a sticky orbit (orange color - S), and a chaotic orbit (red color - C) in our model for a time period of $10^4$ time units. The horizontal, blue, dashed line corresponds to the threshold value $10^{-7}$ which separates regular from chaotic motion. The chaotic orbit needs only about 40 time units in order to cross the threshold, while the sticky orbit on the other hand, requires about 4200 time units so as to reveal its true chaotic nature.}
\label{salis}
\end{figure}

The time-evolution of SALI strongly depends on the nature of the computed orbit. When the orbit is regular the SALI exhibits small fluctuations around non zero values, while in the case of chaotic orbits the SALI, after a small transient period, it tends exponentially to zero approaching the limit of the accuracy of the computer $(10^{-16})$. Therefore, the particular time-evolution of the SALI allow us to distinguish fast and safely between regular and chaotic motion. The time-evolution of a regular (R) and a chaotic (C) orbit, for a time period of $10^4$ time units, is presented in Fig. \ref{salis}. We observe, that both regular and chaotic orbits exhibit the expected behavior. Nevertheless, we have to define a specific numerical threshold value for determining the transition from regularity to chaos. After conducting extensive numerical experiments, integrating many sets of orbits, we conclude that a safe threshold value for the SALI, taking into account the total integration time of $10^4$ time units, is the value $10^{-7}$. The horizontal, blue, dashed line in Fig. \ref{salis} corresponds to that threshold value which separates regular from chaotic motion. In order to decide whether an orbit is regular or chaotic, one may use the usual method according to which we check after a certain and predefined time interval of numerical integration, if the value of SALI has become less than the established threshold value. Therefore, if SALI $< 10^{-7}$ the orbit is chaotic, while if SALI $ > 10^{-4}$ the orbit is regular. Therefore, the distinction between regular and chaotic motion is clear and beyond any doubt when using the SALI method. For the computation of SALI we used the \verb!LP-VI! code \citep{CMD14}, a fully operational routine which efficiently computes a suite of many chaos indicators for dynamical systems in any number of dimensions.

\begin{figure*}[!t]
\centering
\resizebox{0.7\hsize}{!}{\includegraphics{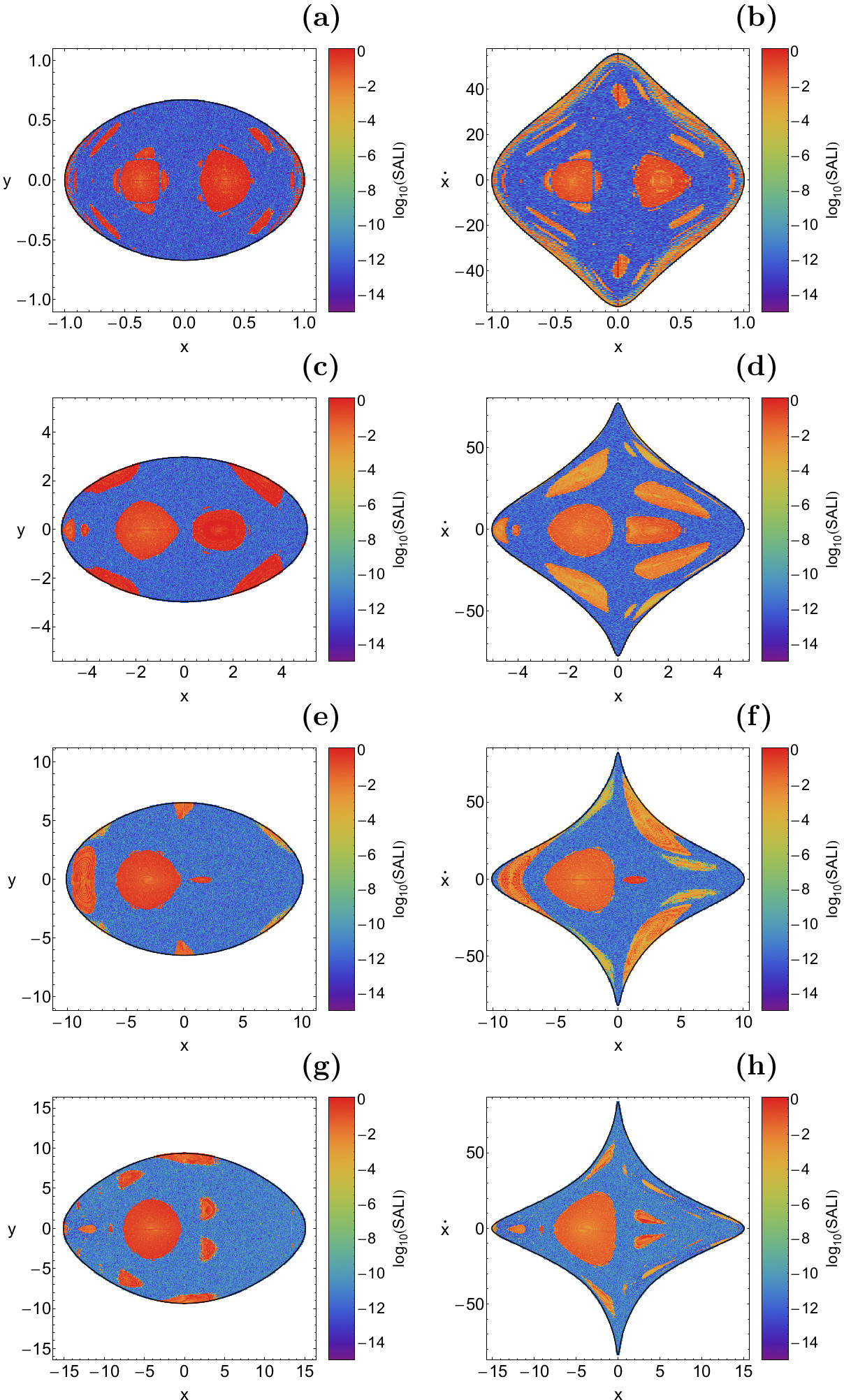}}
\caption{Orbital structure of the (left column): configuration $(x,y)$ space and (right column): phase $(x,\dot{x})$ space of the barred galaxy model for four values of the Jacobi integral $E$. (first row): $E = -2840$, (second row): $E = -1400$, (third row): $E = -1020$ and (fourth row): $E = -895$.}
\label{grids}
\end{figure*}

In our study each orbit was numerically integrated for a time interval of $10^4$ time units. The particular choice of the total integration time is an element of great importance, especially in the case of the so called ``sticky orbits" (i.e., chaotic orbits that behave as regular ones during long periods of time), for which the SALI lies in the interval $[10^{-8},10^{-4}]$. A sticky orbit could be easily misclassified as regular by any chaos indicator\footnote{Generally, dynamical methods are broadly split into two types: (i) those based on the evolution of sets of deviation vectors in order to characterize an orbit and (ii) those based on the frequencies of the orbits which extract information about the nature of motion only through the basic orbital elements without the use of deviation vectors.}, if the total integration interval is too small, so that the orbit do not have enough time in order to reveal its true chaotic character. Thus, all the sets of orbits of a given grid were integrated, as we already said, for $10^4$ time units. All the sticky orbits which do not show any signs of chaoticity after $10^4$ time units are counted as regular ones, since that vast sticky periods are completely out of scope of our research.

\section{Numerical results}
\label{numres}

In order to explore how the value of the energy affects the overall orbital structure of our barred galaxy model, we use the normal procedure according to which we let the energy vary, while fixing the values of all the other parameters of our galactic models, according to SM. We note that the particular value of the energy determines the maximum possible value of the $x$ coordinate $(x_{\rm max})$. To select the energy levels, we chose those values of the energy which give $x_{\rm max} = \{1,5,10,15\}$. Fig. \ref{grids}(a-h) shows grids of initial conditions of orbits on both the configuration and the phase space that we have classified for four values of the energy integral $E$. For all the initial conditions the values of the SALI are plotted using different colors according to the color bar. The stability regions of regularity are indicated by light reddish colors, while the chaotic domains by dark blue/purple colors. All intermediate colors correspond to sticky orbits. The outermost black solid line in the configuration space is the Zero Velocity Curve (ZVC) which is defined as $\Phi_{\rm eff}(x,y) = E$, while in the phase space is the limiting curve given by
\begin{equation}
f(x,\dot{x}) = \frac{1}{2} \dot{x}^2 + \Phi_{\rm eff}(x, y = 0) = E.
\label{zvc}
\end{equation}
In all cases we observe the presence of several stability islands which correspond to different resonant orbital families, while all the regular regions are surrounded by a unified chaotic sea. Near the center of the two types of the grids there are the stability islands which correspond to the main 1:1 loop orbits. At the left part, with $x_0 < 0$, we have the retrograde 1:1 resonant orbits, while on the right part, with $x_0 > 0$, is the location of the prograde 1:1 resonant orbits. It is seen that as we proceed to higher energy levels the area of the prograde 1:1 orbits is constantly reduced and for $E = -895$ there is no indication of the corresponding stability island. Furthermore, one can observe in the phase planes that with increasing energy there is a growth in the allowed velocity $\dot{x}$ of the stars near the central region of the barred galaxy.

\begin{figure*}[!t]
\centering
\resizebox{\hsize}{!}{\includegraphics{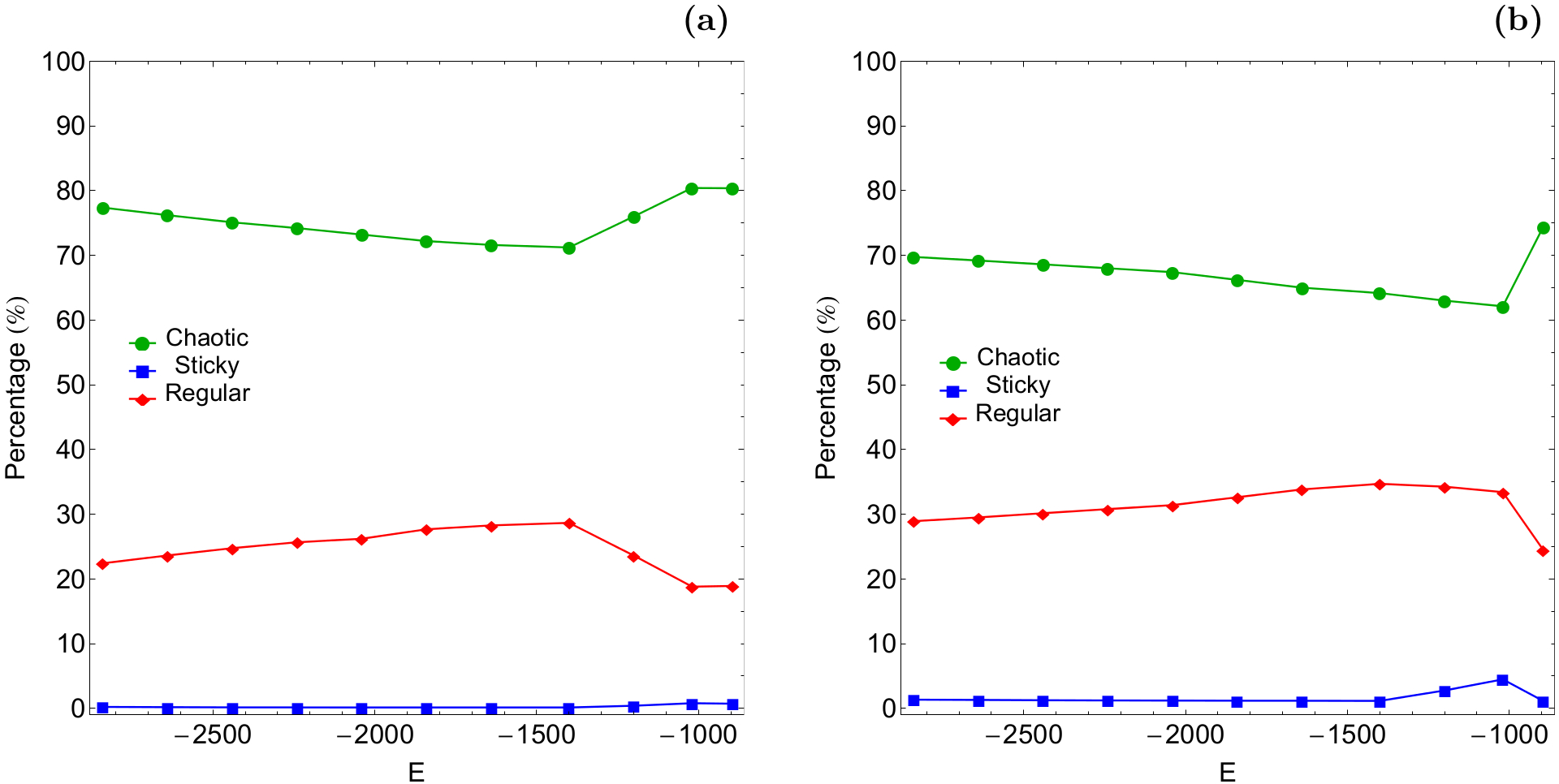}}
\caption{Evolution of the percentages of regular, sticky and chaotic orbits on the (a-left): configuration space and (b-right): phase space, when varying the total orbital energy $E$.}
\label{percs}
\end{figure*}

It would be very interesting to monitor the evolution of the percentages of regular, sticky and chaotic orbits as the value of the total orbital energy varies. The resulting percentages are shown in Fig. \ref{percs}(a-b). Regular orbits are those with SALI $ > 10^{-4}$, sticky orbits are those with $10^{-4} \leq $ SALI $ \leq 10^{-7}$, while chaotic orbits are those with SALI $< 10^{-7}$ after $10^4$ time units of numerical integration. Fig. \ref{percs}a corresponds to the configuration $(x,y)$ space. It is evident that in the interval $-2900 < E < -1400$ there is a monotone evolution of the percentages; the rate of regular orbits exhibit a minor increase, while the percentage of chaotic orbits slightly decreases. For larger values of the energy $(E > -1400)$ the patterns are reversed and for $E > -1000$ the percentages seem to saturate around 80\% for the chaotic orbits and around 20\% for the regular orbits. Things are quite similar in the phase space as we can see in panel (b) of Fig. \ref{percs}. Indeed the same monotone behaviour is observed, while the turning point appears later (at about $E = -1000$) with respect to what we seen for the configuration space. In both cases the percentage of sticky orbits holds always very low values (lower than 5\%).

The color-coded grids in the configuration $(x,y)$ as well as in the phase $(x,\dot{x})$ space provide sufficient information on the phase space mixing however, for only a fixed value of the total orbital energy. H\'{e}non back in the late 60s \citep{H69}, introduced a new type of plane which can provide information about stability and chaotic regions using the section $y = \dot{x} = 0$, $\dot{y} > 0$. In other words, all the orbits of the stars of the barred galaxy are launched from the $x$-axis with $x = x_0$, parallel to the $y$-axis $(y = 0)$. Consequently, in contrast to the previously discussed types of planes, only orbits with pericenters on the $x$-axis are included and therefore, the value of the energy $E$ can be used as an ordinate. In this way, we can monitor how the energy influences the overall orbital structure of our barred galaxy model using a continuous spectrum of energy values rather than few discrete energy levels.

\begin{figure}[!t]
\centering
\includegraphics[width=\hsize]{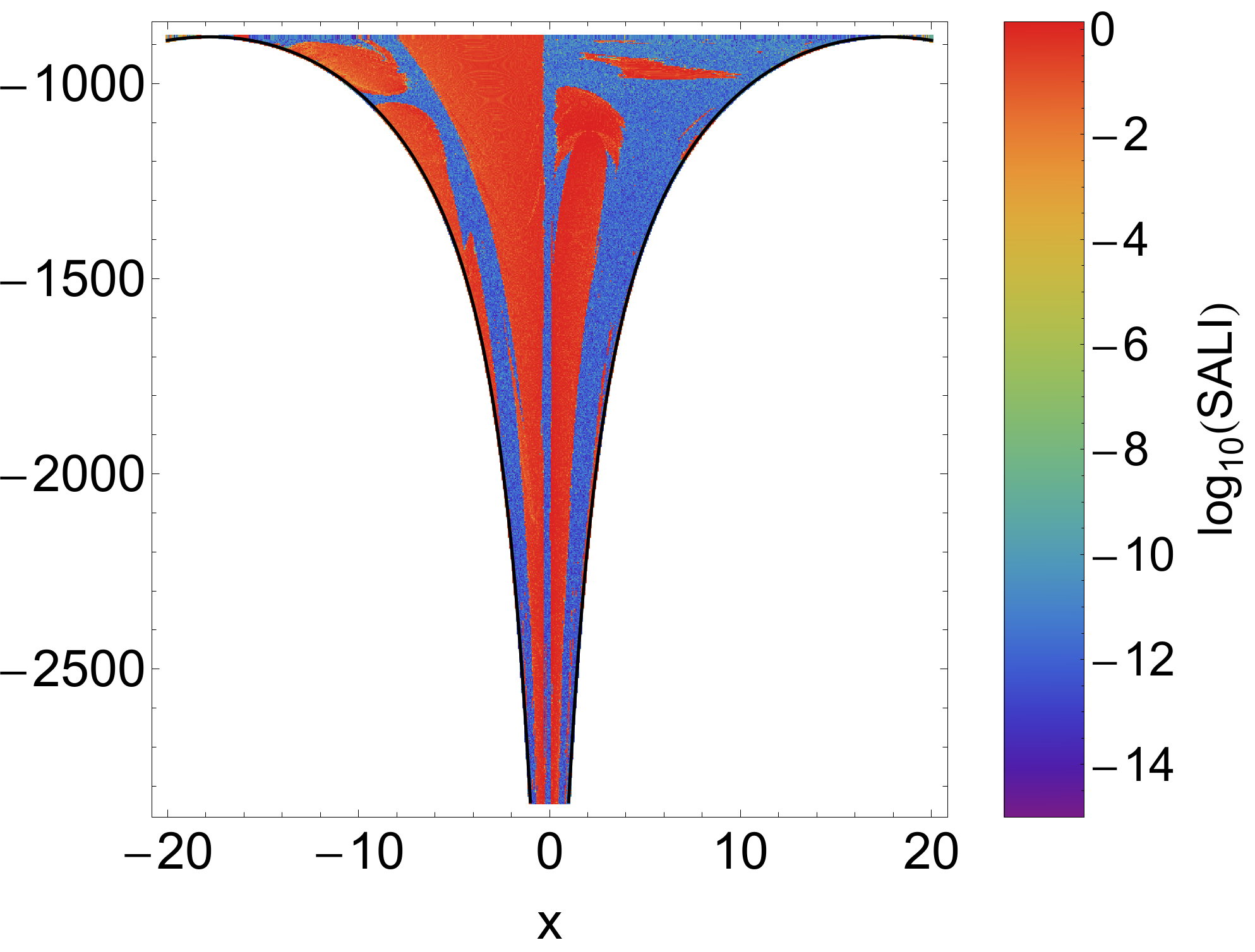}
\caption{Orbital structure of the $(x,E)$ space, when $E \in [-2840,-880]$.}
\label{xE}
\end{figure}

\begin{figure}[!t]
\centering
\includegraphics[width=\hsize]{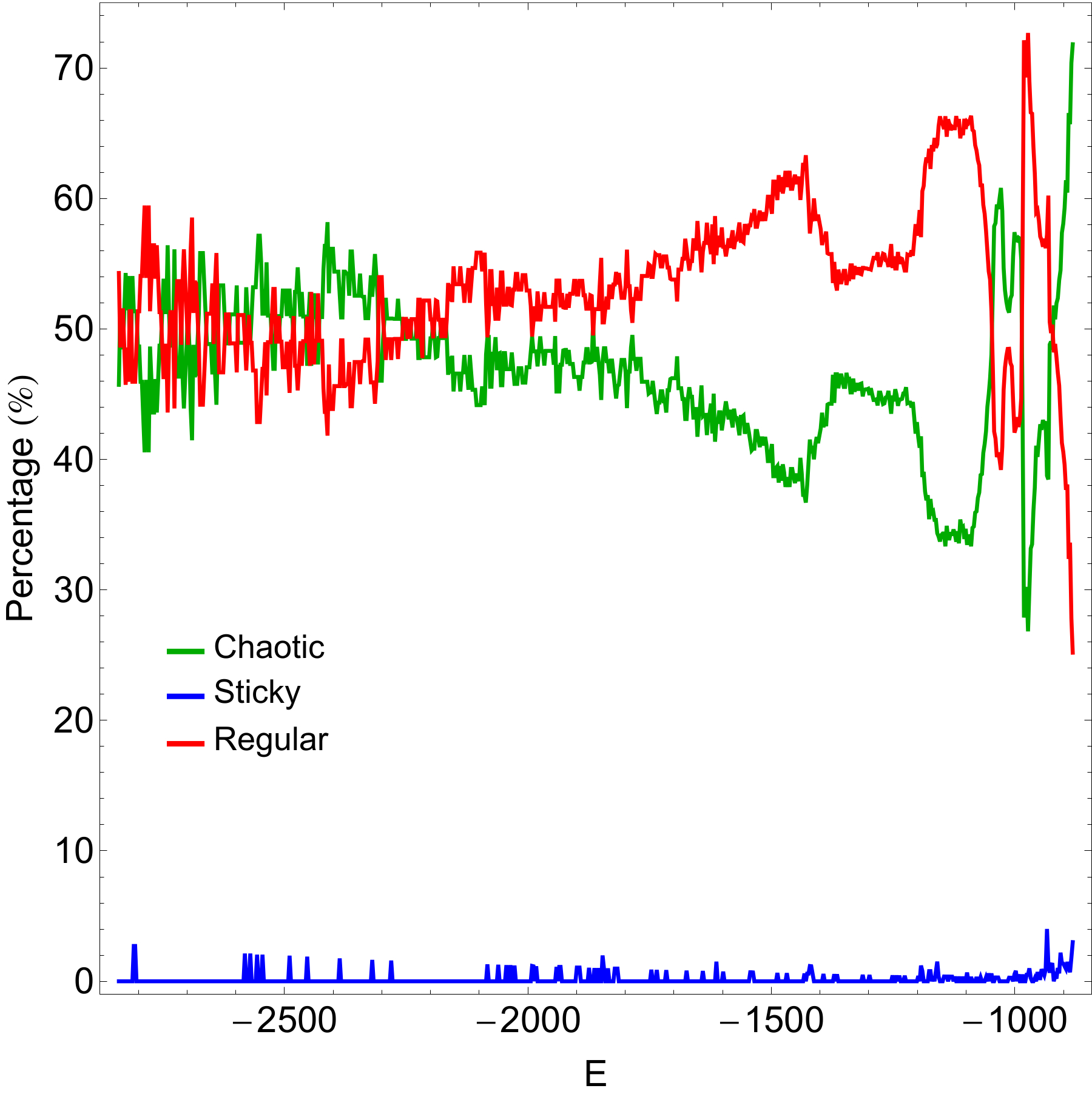}
\caption{Evolution of the percentages of regular, sticky and chaotic orbits on the $(x,E)$ space.}
\label{pxE}
\end{figure}

In Fig. \ref{xE} we present the orbital structure of the $(x,E)$ plane when $E \in [-2840,-880]$. Here the evolution of the prograde 1:1 stability island is better explained, as we can see how it breaks for relatively high energy levels. The evolution of the percentages of regular, sticky and chaotic orbits on the $(x,E)$ plane is shown in Fig. \ref{pxE}. It is seen that for about $-2840 < E < -1900$ the rates of both regular and chaotic orbits evolve almost identically as they fluctuate around 50\%. For larger values of the energy the rates start to diverge thus following mirror-imaged patterns. In particular, in the interval $-1900 < E < -1100$ the rate of regular orbits is always higher than that of the chaotic orbits, while for $E > -1100$ the patterns are interchangeable. Once more, sticky orbits possess very low rates as our computations reveal that the corresponding value is always less than 5\% of the total tested initial conditions of orbits per energy value.

Before closing this Section we would like to emphasize that when modelling the kinematics derived from observations, it is important to know the proportion of low and high energy stars, compared to overall stars. This information however is mainly obtained through $N$-body simulations, where the initial conditions of the orbits of the stars are generated using a distribution function of the velocities. Then it is rather easy to determine the relative fraction of low and high energy stars. In our case on the other hand, we do not use any type of distribution function, while every set of initial conditions of orbits corresponds to a particular level of the total orbital energy (see e.g., Fig. \ref{grids}). On this basis, it is not possible to have an estimation of fraction of low and high energy stars.

\section{Conclusions and discussion}
\label{disc}

In this paper we tried to distinguish between ordered and chaotic motion in a simple two dimensional barred galaxy model. Apart from the bar the galaxy contains a central, spherically symmetric, dense nucleus and a flat disk. Our aim was to investigate how the total orbital energy influences the level of chaos in our barred galaxy model. Our results strongly suggest, that the level of chaos is indeed very dependent on the value of the energy.

We defined for several values of the energy integral, dense uniform grids of $1024 \times 1024$ initial conditions in several types of two dimensional planes and then we identified regions of order and chaos. For the numerical integration of the orbits in each type of grid, we needed about between 1.5 hour and 12 days of CPU time on a Pentium Dual-Core 2.2 GHz PC. For each initial condition, the maximum time of the numerical integration was set to be equal to $10^4$ time units.

It is generally accepted, that a barred galaxy with a central spherical nucleus and a disk is surely a very complex stellar entity and therefore, we need to assume some necessary simplifications and assumptions in order to be able to study numerically the orbital behavior of such complicated stellar system. For this purpose, our model is intentionally simple and contrived, in order to give us the ability to study all the different aspects, regarding the kinematics and dynamics of the model. Nevertheless, contrived models can provide an insight into more realistic stellar systems, which unfortunately are very difficult to be studied, if we take into account all the astrophysical aspects (i.e. gas, spirals, mergers, etc). Self-consistent models on the other hand, are mainly used when conducting $N$-body simulations. However, this application is entirely out of the scope of the present paper. Once again, we have to point out that the simplicity of our model is necessary, otherwise it would be extremely difficult, or even impossible, to apply the detailed numerical calculations presented in this study. Similar galaxy models with the same limitations and assumptions were used successfully, several times in the past, for investigating the orbital structure in much more complicated galactic systems \citep[e.g.,][]{CI09,C12,Z12b,Z13}.

We consider the present results as an initial effort and also a promising step in the task of understanding the orbital structure of barred galaxies. Taking into account that our outcomes are encouraging, it is in our future plans to properly modify our dynamical model thus expanding our investigation into three dimensions and exploring how the basic parameters influence the nature of the three dimensional (3D) orbits. Furthermore, it would be very illuminating if we could classify the regular orbits into different resonant families.

\section*{Acknowledgments}

The author would like to express his warmest thanks to Dr. Rain Kipper for the careful reading of the manuscript and for all the apt suggestions and comments which allowed us to improve both the quality and the clarity of the paper.


\begin{thebibliography}{}
\footnotesize

\bibitem[\protect\citeauthoryear{Athanassoula}{1984}]{A84} Athanassoula, E. 1984, Phys. Rep., 114, 319

\bibitem[\protect\citeauthoryear{Athanassoula}{1992}]{A92} Athanassoula, E. 1992, MNRAS, 259, 345

\bibitem[\protect\citeauthoryear{Athanassoula et al.}{1983}]{ABMP83} Athanassoula, E., Bienayme, O., Martinet, L., Pfenniger, D. 1983, A\&A, 127, 349

\bibitem[\protect\citeauthoryear{Binney \& Tremaine}{2008}]{BT08} Binney, J., Tremaine, S. 2008, Galactic Dynamics, Princeton Univ. Press, Princeton, USA

\bibitem[\protect\citeauthoryear{Bountis et al.}{2012}]{BMA12} Bountis, T., Manos, T., Antonopoulos, Ch. 2012, CeMDA, 113, 63

\bibitem[\protect\citeauthoryear{Caranicolas}{2012}]{C12} Caranicolas, N.D. 2012, MNRAS, 423, 2668

\bibitem[\protect\citeauthoryear{Caranicolas \& Innanen}{2009}]{CI09} Caranicolas, N.D., Innanen, K.A. 2009, Astronomische Nachrichten, 330, 20

\bibitem[\protect\citeauthoryear{Caranicolas \& Papadopoulos}{2004}]{CP04} Caranicolas, N.D., Papadopoulos, N.J. 2004, Astronomical and Astrophysical Transactions, 23, 229

\bibitem[\protect\citeauthoryear{Caranicolas \& Papadopoulos}{2007}]{CP07} Caranicolas, N.D., Papadopoulos, N.J. 2007, Astronomische Nachrichten, 328, 556

\bibitem[\protect\citeauthoryear{Caranicolas \& Zotos}{2010}]{CZ10} Caranicolas, N.D., Zotos, E.E. 2010, New Astronomy, 15, 427

\bibitem[\protect\citeauthoryear{Carpintero et al.}{2014}]{CMD14} Carpintero D.D., Maffione N., Darriba L. 2014, Astronomy and Computing, 5, 19

\bibitem[\protect\citeauthoryear{Combes et al.}{1990}]{CDFP90} Combes, F., Debbasch, F., Friedli, D., Pfenniger, D. 1990, A\&A, 233, 82	

\bibitem[\protect\citeauthoryear{Contopoulos \& Grosb{\o}l}{1989}]{CG89} Contopoulos, G., Grosb{\o}ol, P. 1989, A\&AR, 1, 261

\bibitem[\protect\citeauthoryear{Ernst \& Peters}{2014}]{EP14} Ernst, A., Peters, T. 2014, MNRAS, 443, 2579

\bibitem[\protect\citeauthoryear{Eskridge et al.}{2000}]{Ee00} Eskridge, P.B., Frogel, J.A., Pogge, R.W., Quillen, A.C., Davies, R.L., DePoy, D.L., Houdashelt, M.L., et al. 2000, AJ, 119, 356

\bibitem[\protect\citeauthoryear{Hasan \& Norman}{1990}]{HN90} Hasan, H., Norman, C.A. 1990, ApJ, 361, 69

\bibitem[\protect\citeauthoryear{Hasan et al.}{1993}]{HPN93} Hasan, H., Pfenniger, D., Norman, C. 1993, ApJ, 409, 91

\bibitem[\protect\citeauthoryear{H\'{e}non}{1969}]{H69} H\'{e}non, M. 1969, A\&A, 1, 223

\bibitem[\protect\citeauthoryear{Jung \& Zotos}{2015}]{JZ15} Jung, Ch., Zotos, E.E. 2015, PASA, 32, id.e042

\bibitem[\protect\citeauthoryear{Jung \& Zotos}{2016}]{JZ16} Jung, Ch., Zotos, E.E. 2016, MNRAS, 3, 2583

\bibitem[\protect\citeauthoryear{Kaufmann \& Contopoulos}{1996}]{KC96} Kaufmann, D.E., Contopoulos, G. 1996, A\&A, 309, 381

\bibitem[\protect\citeauthoryear{Kaufmann \& Patsis}{2005}]{KP05} Kaufmann, D., Patsis, P. 2005, ApJ, 624, 693

\bibitem[\protect\citeauthoryear{Kormendy \& Kennicutt}{2004}]{KK04} Kormendy, J., Kennicutt, R.C., Jr. ARA\&A, 42, 603

\bibitem[\protect\citeauthoryear{Manos \& Athanassoula}{2011}]{MA11} Manos, T., Athansssoula, E. 2011, MNRAS, 415, 629

\bibitem[\protect\citeauthoryear{Manos et al.}{2013}]{MBS13} Manos, T., Bountis, T., Skokos, Ch. 2013, J. Phys. A: Math.

\bibitem[\protect\citeauthoryear{Oll\'{e} \& Pfenniger}{1998}]{OP98} Oll\'{e}, M., Pfenniger, D. 1998, A\&A, 334, 829

\bibitem[\protect\citeauthoryear{Pfenniger}{1984}]{P84} Pfenniger, D. 1984, A\&A, 134, 373

\bibitem[\protect\citeauthoryear{Pfenniger}{1996}]{P96} Pfenniger, D., 1996, in Buta R., Crocker D. A., Elmegreen B. G. eds, ASP Conf. Ser. Vol. 91, Barred Galaxies. Astron. Soc. Pac., San Francisco, p. 273

\bibitem[\protect\citeauthoryear{Pichardo et al.}{2004}]{PMM04} Pichardo, B., Martos, M., Moreno, E. 2004, ApJ, 609, 144

\bibitem[\protect\citeauthoryear{Press et al.}{1992}]{PTVF92} Press H.P., Teukolsky S.A, Vetterling W.T., Flannery B.P. 1992, Numerical Recipes in FORTRAN 77, 2nd Ed., Cambridge Univ. Press, Cambridge, USA

\bibitem[\protect\citeauthoryear{Sellwood \& Wilkinson}{1993}]{SW93} Sellwood, J., Wilkinson, A. 1993, Rep. Prog. Phys., 56, 173

\bibitem[\protect\citeauthoryear{Sheth et al.}{2003}]{SRSS03} Sheth, K., Regan, M.W., Scoville, N.Z., Strubbe, L.E. 2003,

\bibitem[\protect\citeauthoryear{Skokos}{2001}]{S01} Skokos C. 2001, J. Phys. A: Math. Gen., 34, 10029

\bibitem[\protect\citeauthoryear{Skokos et al.}{2002a}]{SPA02a} Skokos, Ch., Patsis, P.A., Athanassoula, E. 2002a MNRAS, 333 847

\bibitem[\protect\citeauthoryear{Skokos et al.}{2002b}]{SPA02b} Skokos, Ch., Patsis, P.A., Athanassoula, E. 2002n, MNRAS, 333, 861

\bibitem[\protect\citeauthoryear{Zotos}{2012a}]{Z12a} Zotos, E.E. 2012a, New Astronomy, 17, 576

\bibitem[\protect\citeauthoryear{Zotos}{2012b}]{Z12b} Zotos, E.E. 2012b, ApJ, 750, 56

\bibitem[\protect\citeauthoryear{Zotos}{2012c}]{Z12c} Zotos, E.E. 2012c, RAA, 12, 500

\bibitem[\protect\citeauthoryear{Zotos}{2013}]{Z13} Zotos, E.E. 2013, PASA, 30, 12

\bibitem[\protect\citeauthoryear{Zotos \& Carpintero}{2013}]{ZC13} Zotos, E.E., Carpintero, D.D. 2013, CeMDA, 116, 417

\end{thebibliography}
\end{document}